\documentclass[aip, apl, showpacs, reprint, preprintnumbers,amsmath,amssymb,superscriptaddress,floatfix]{revtex4-1}

\usepackage[]{graphicx}      
\usepackage{bm}             	
\usepackage{times}			
\usepackage{tabularx}
\usepackage{color}
\usepackage{dcolumn}		
\usepackage{amsmath}
\usepackage{amssymb}
\usepackage{amsfonts}
\usepackage{times}
\usepackage{tabularx}
\usepackage{xcolor,colortbl}

\usepackage{amsmath}

\makeatletter
\newcommand\xleftrightarrow[2][]{%
  \ext@arrow 9999{\longleftrightarrowfill@}{#1}{#2}}
\newcommand\longleftrightarrowfill@{%
  \arrowfill@\leftarrow\relbar\rightarrow}
\makeatother

\begin{document}
\title{Effect of Tantalum spacer thickness and deposition conditions on the properties of MgO/CoFeB/Ta/CoFeB/MgO free layers}
\author{T. Devolder}
\email{thibaut.devolder@u-psud.fr}
\affiliation{Centre de Nanosciences et de Nanotechnologies, CNRS, Univ. Paris-Sud, Universit\'e Paris-Saclay, 91120 Palaiseau, France}
\author{S. Couet}
\affiliation{imec, Kapeldreef 75, 3001 Heverlee, Belgium}
\author{J. Swerts}
\author{S. Mertens}
\author{S. Rao}
\author{G. S. Kar}
\affiliation{imec, Kapeldreef 75, 3001 Heverlee, Belgium}

\date{\today}                                           
%
%
\begin{abstract}
To get stable perpendicularly magnetized tunnel junctions at small device dimensions, composite free layers that comprise two MgO/FeCoB interfaces as sources of interface anisotropy are generally used. Proper cristallisation and annealing robustness is typically ensured by the insertion of a spacer layer of the early transition metal series within the FeCoB layer. We study the influence of the spacer thickness and growth condition on the switching metrics of tunnel junctions thermally annealed at 400$^\circ$C for the case of 1-4 \r{A} Ta spacers. Thick Ta spacer results in a large anisotropies indicative of a better defined top FeCoB/MgO interface, but this is achieved at the systematic expense of a stronger damping. For the best anisotropy-damping compromise, junctions of diameter 22 nm can still be stable and spin-torque switched. Coercivity and inhomogeneous linewidth broadening, likely arising from roughness at the FeCoB/Ta interface, can be reduced if a sacrificial Mg layer is inserted before the Ta spacer deposition. 
\end{abstract}

\maketitle

%
%

The design of magnetic tunnel junctions (MTJ) suitable for spintronic applications involves material science and nanomagnetism aspects. An emblematic example is the spin-transfer-torque magnetic random access memory (STT-MRAM) technology \cite{Khvalkovskiy_basic_2013} which requires to include in a single nanodevice the physics of perpendicular magnetic anisotropy (PMA), tunnel magneto-resistance (TMR), interlayer exchange coupling, microwave magnetization dynamics, alloy metallurgy, as well as crystallization and elemental diffusion kinetics \cite{worledge_spin_2011, gajek_spin_2012, devolder_material_2018}. While when studied on model systems, these phenomena are understood and the corresponding technique are well mastered, their quantitative prediction in complex systems remains challenging.
This is especially true when the layer thicknesses are in the near-monolayer regime when the concepts based on diffusive transport (such as spin pumping induced damping) are out of their range of validity while ab-initio approaches can't be applied because of a insufficient knowledge on the sample structure. For instance, it still difficult to predict or understand quantitatively the consequences of the insertion of a metallic spacer at the middle of a magnetic layer despite the fact that this material science trick is commonly practiced \cite{sato_properties_2014, thomas_quantifying_2015, kim_ultrathin_2015, liu_high_2016} in the free layers of STT-MRAMs. 

In this paper, we study experimentally the influence of a Tantalum spacer layer inserted within the dual MgO FeCoB-based free layer of optimized MTJs with PMA. We measure the transport properties as well as the magnetic anisotropy and the Gilbert damping of samples having faced the standard 400$^\circ$C annealing of a CMOS back end of line process.
We show that both the damping and the magneto-crystalline anisotropy increase with the Ta thickness. We also show that the deposition condition (with or without a Mg sacrificial layer before the deposition of the spacer) influences the homogeneity of the magnetic properties. 

\begin{figure}
\hspace*{-1cm}\includegraphics[width=10.5 cm]{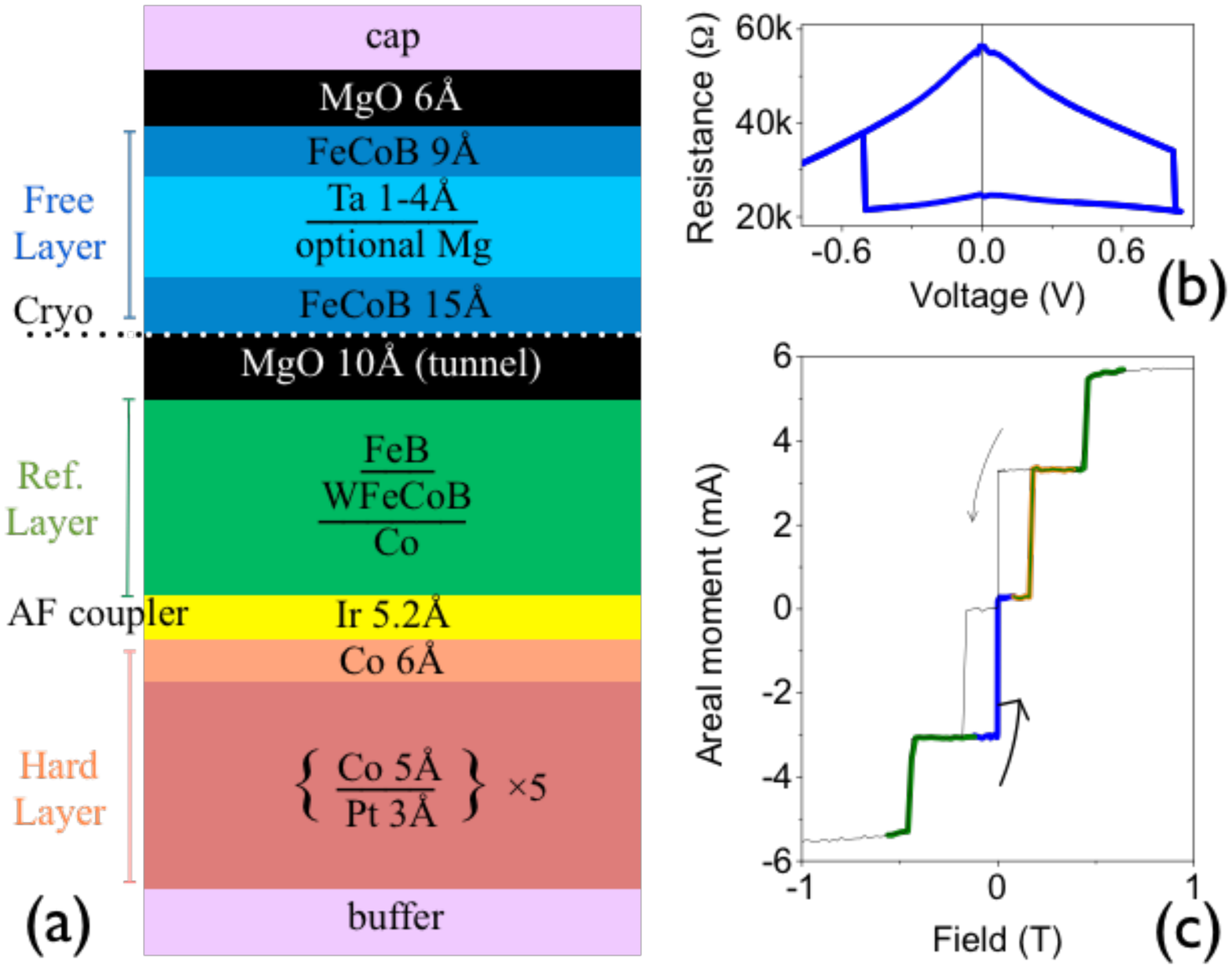}{\centering}
\caption{(a). Sketch of the sample stack. (b) Hysteresis loop of resistance versus voltage measured on a circular device of diameter 22 nm in which the free layer spacer is Mg (6.5) / Ta (3). (c) Hysteresis loop of the areal magnetic moment. The arrows indicate the field sweeping directions. The color are chosen to emphasize the layers that are switching. }
\label{STACK}
\end{figure}

Our objective is to understand the influence of the Tantalum spacer onto the properties of free layers of perpendicularly magnetized MTJ that have undergone 400$^\circ$C annealing and in which element interdiffusion is minimized by a proper engineering of the Boron content \cite{devolder_material_2018} within the FeCoB parts of the free layer. We consider free layers embodied in state-of-the-art bottom-pinned MTJs. For this study, we consider the generic stack sketched in Fig.~\ref{STACK}(a). It is deposited by physical vapor deposition and then annealed at 400$^\circ$C for 30 min in a 1 T perpendicular magnetic field. The layer compositions are described from bottom to top with the numbers denoting the thickness in \r{A}.  

The stack organisation is: Hard Layer / Ir (5.2) / Reference Layer / MgO (rf) / Free layer / MgO (rf) / cap. Following previous optimizations \cite{couet_impact_2017, devolder_material_2018}, the Hard Layer is the conventional [Co (5) / Pt (3) ]$_{\times 5}$ / Co (6). Antiferromagnetic coupling with the reference layer is supplied by the iridium spacer. Note that the hysteresis loops comprise 4 steps for each sweeping direction [see Fig.~\ref{STACK}(c)]. This contrasts with anterior configurations using the weaker antiferro-couplers (e.g. Ru) which allowed the RL and HL to switch separately, resulting in loops comprising only 3 successive steps (RL, FL and then finally HL switching) \cite{devolder_performance_2013, devolder_evolution_2016, devolder_annealing_2017, couet_impact_2017}.  The present reference layer is Co (6) / WFeCoB (8) / FeB (9). \textcolor{black}{We have chosen WFeCoB to benefit from a strong interlayer exchange coupling with the cobalt layer and because it induces a large anisotropy \cite{couet_impact_2017} in the neighboring Fe(Co)B layers. We use FeB to benefit from strong interface anisotropy with MgO \cite{kubota_enhancement_2012, konoto_effect_2013}. As clear from Fig. 1b, these two features prevent the dynamical back-hopping of the antiparallel-to-Parallel transition observed formerly with less stiff reference layers \cite{devolder_time-resolved_2016, devolder_evolution_2016}. The properties of a very similar reference layer are detailed in ref.~\onlinecite{devolder_offset_2019}}.

While the hard and the reference layer were deposited at room temperature, the deposition of the tunnel oxide is followed by a sample refrigeration at cryogenic temperatures (80 K) which allows a better wetting of the subsequently deposited FeCoB on the MgO tunnel oxide, with benefits\cite{swerts_cryogenic_nodate} (+20\%) in TMR and RA. The free layer is then Fe$_{52.5}$Co$_{17.5}$B$_{30}$ (15) / spacer / Fe$_{52.5}$Co$_{17.5}$B$_{30}$ (9). 8 variants of the spacer were used, consisting of Tantalum thickness of 1, 2, 3 or 4 \r{A} deposited on an optional sacrificial Mg (6.5) layer \cite{swerts_beol_2015}. Note that thicker Ta spacers lead generally to a loss of the exchange coupling between two adjacent FeCoB layers \cite{le_goff_effect_2014}. 

%

We studied our MTJs by vibrating sample magnetometry (VSM), current-in-plane tunneling (CIPT) and Vector Network Ferromagnetic resonance (VNA-FMR \cite{bilzer_vector_2007}) in out-of-plane fields. VSM minor loops (not shown) of the free layer are square except for the 1 \r{A} thick Tantalum spacers for which the easy magnetization axes are in-plane, as will be confirmed by VNA-FMR.
CIPT is performed for out-of-plane applied fields of $\mu_0 H_z = \pm 150~\textrm{mT}$. While this is suitable for systems with perpendicular easy axes or weak easy-plane anisotropy, this field is insufficient to fully saturate the sample with a Mg/Ta(1 \r{A}) spacer. For a fair comparison of the TMR among the samples, the TMR measured on the latter sample at 150 mT are renormalized by the factor $H_{k}^\textrm{eff} /H_z$ where the effective anisotropy field $H_k-M_s$ will be extracted by VNA-FMR.

\begin{figure}
\includegraphics[width=8.5 cm]{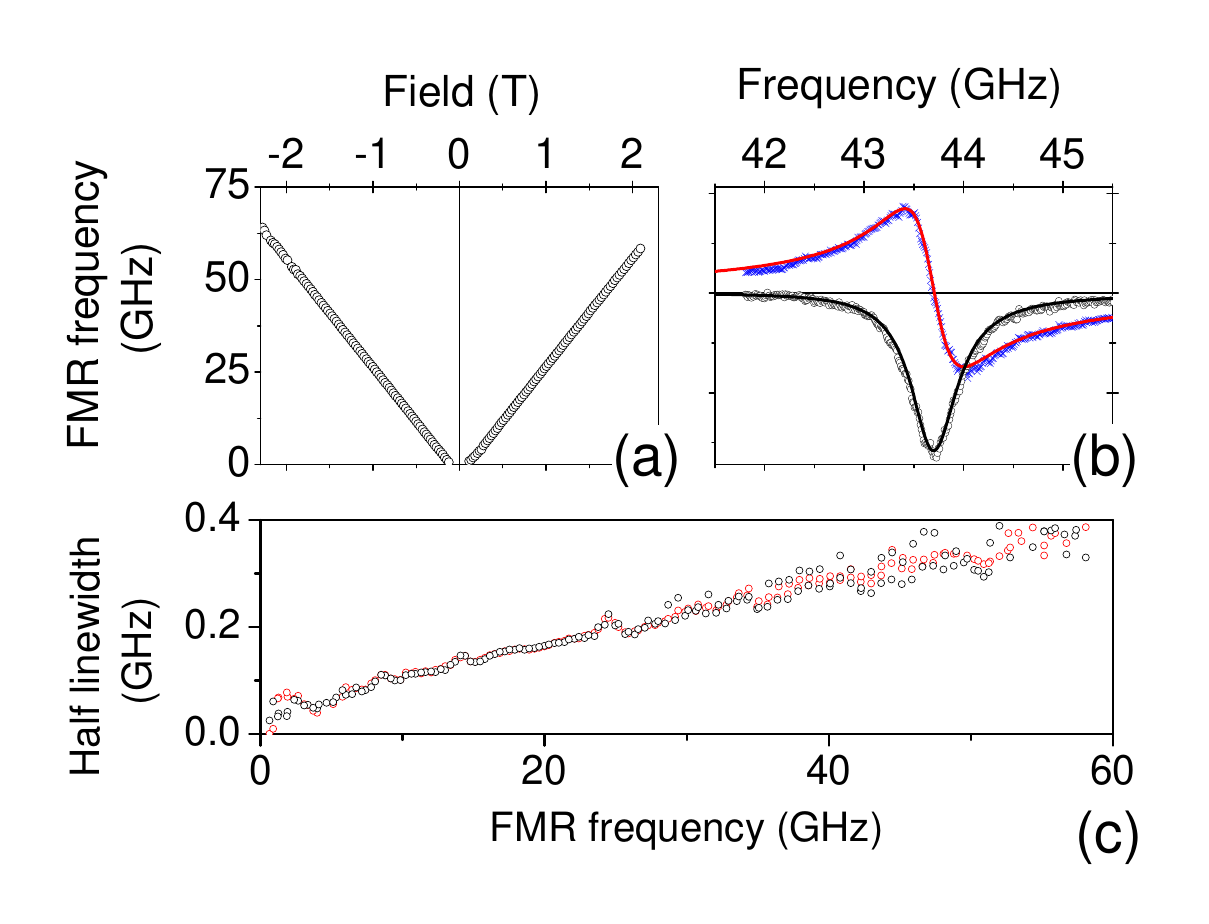}
\caption{(Color online). Ferromagnetic resonance properties of the free layer in which the spacer is Ta 1 \r{A}. (a) FMR frequency versus out-of-plane applied field. (b) Real and imaginary parts of the free layer permeability recorded (symbols) in a field of 1.6 Tesla. The lines are fits with macrospin expressions, leading modeled linewidths of respectively 592 (real part) and 563 MHz (imaginary part). (c) Modeled linewidth versus FMR frequency as extracted from the real part of the permeability (red) or the imaginary part (black). The average slope is $0.006 \pm 0.0002$.}
\label{FMR}
\end{figure}

VNA-FMR is used for an in-depth characterization of the free layer properties. The joined field and frequency resolutions of VNA-FMR is convenient to study the properties of each subsystem of the MTJ \cite{devolder_evolution_2016}. Indeed the free layer FMR frequency versus field curve inverses its slope at the free layer coercivity (a couple of mT), while the fixed system eigenmodes undergo frequency jumps or slope inversions at the characteristic fields of the fixed system hysteresis loop that can be clearly identified by the VSM characterizations. The analysis of the free layer FMR (Fig.~\ref{FMR}) is conducted the following way. For each applied field, VNA-FMR provides the real and imaginary parts of the permeability. They are fitted separated [Fig.~\ref{FMR}(b)] with the adequate macrospin expressions \cite{devolder_using_2017} to provide the FMR frequencies and their frequency linewidth. The FMR frequencies extracted from the real parts and from the imaginary parts generally agree within a couple of MHz. For fields larger than the FL coercivity the FMR frequencies are independent of the field history and are modeled with the expression $\frac{\gamma_0}{2\pi} (|H_z| + H_k-M_s)$ where the gyromagnetic factor $\gamma_0$ and the effective anisotropy fields $H_k-M_s$ are the fitted parameters [Fig.~\ref{FMR}(b)]. 

The FMR linewidth $\frac{1}{2\pi} \Delta \omega_{\textrm{perp}}$ extracted from the real and imaginary parts of the permeability can differ more substantially [Fig.~\ref{FMR}(c)]. Our modest signal-to-noise ratio as well as impedance mismatch artefacts than can't be accounted for perfectly can result in differences in linewidth of up to 10\% in the worst cases [see such example in Fig.~\ref{FMR}(b)]. Despite this uncertainty we can reliably model the linewidth using  $\Delta \omega_{\textrm{perp}} =2 \alpha \omega_{\textrm{perp}} +  \gamma_0 \Delta H_0$ to separate the Gilbert damping $\alpha$ and the samples' inhomogenities contributions $\Delta H_0$ to the total linewidth. The damping is obtained with a precision always better than 0.0005.

%

\begin{figure}
\includegraphics[width=9 cm]{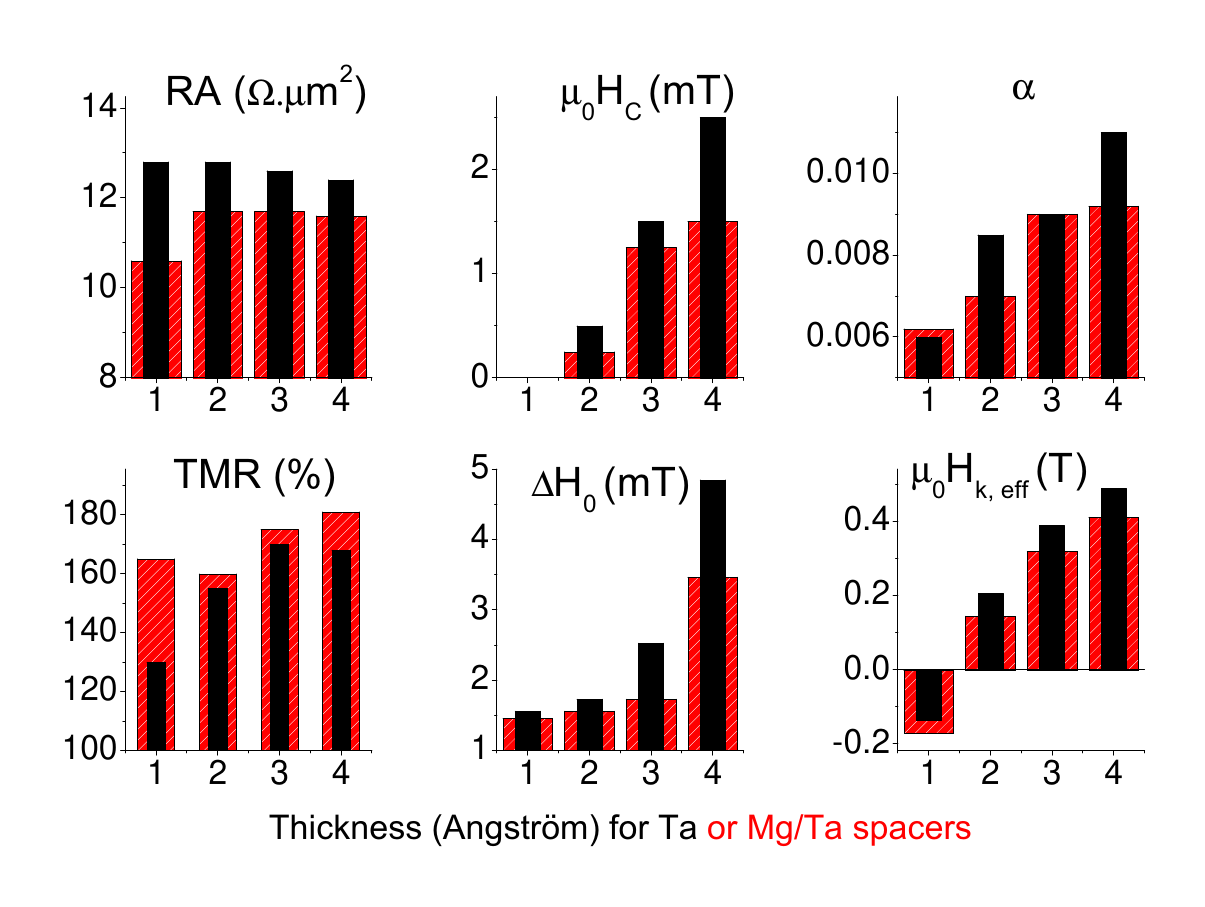}
\caption{(Color online). Transport metrics (resistance area product, tunnel magneto-resistance), magnetometry metrics (coercivity and effective anisotropy field) and ferromagnetic resonance metrics (Gilbert damping and inhomogeneous linewidth broadening) of the different free layer configurations. The shaded wide red bars are for Mg/Ta spacers while the narrow black bars are for solitary Tantalum spacers.}
\label{METRICS}
\end{figure}

The metrics extracted from CIPT and VNAFMR experiments are plotted in Fig.~\ref{METRICS}. Let us first depict the effect of the tantalum thickness. 
The major effect of Ta is to enable a larger effective anisotropy after the annealing: 1 \r{A} is insufficient for the PMA to survive the 400$^\circ$C annealing. The corresponding effective anisotropy stays close to that of a 24 \r{A} FeCoB layer that would comprise a single FeCoB/MgO interface. Several scenario can account for this reduced anisotropy. The top FeCoB free layer may not be in epitaxial relationship with the top MgO because the Ta spacer would not be thick enough to relax the lattice misorientations between the two FeCoB parts of the free layer, or the Ta spacer may not be thick enough to absorb the boron that would then diffuse preferentially to the closest MgO interface, thereby degrading the interface anisotropy and the TMR. 
Fortunately, thicker Ta spacers succeed in maintaining PMA after the 400$^\circ$C annealing, indicating that the top FeCoB/MgO interface now contributes substantially to the overall anisotropy, especially for the thickest Ta spacers for which the FeCoB/MgO interface anisotropy seems fully developed as the effective anisotropy stops increasing further (not shown). The increase of anisotropy with Ta thickness comes unfortunately with a substantial increase of the free layer damping (Fig.~\ref{METRICS}, right column) such that a trade-off has to be done to get high enough anisotropy and low enough damping. Junctions fabricated with the Mg/Ta(3 \r{A} spacers are for instance both stable at diameters as low as 22 nm and can still be switched by spin-torque [fig.~\ref{STACK}(b)]. The lowest damping (0.006) is achieved on the Mg/Ta (1 \r{A}) and Ta (1 \r{A}) samples, and it increases by typically 0.001 per extra \r{A} of Ta spacer. Note also that when passing from annealing at 300$^\circ$C (see ref.~\onlinecite{devolder_material_2018}) to 400$^\circ$C (present case), the damping degraded by typically 0.002. This correlation between high annealing temperature and high damping, as well as the correlation between a large number of Ta atoms within the free layer and a large damping recalls the fact that the damping in FeCo system scales with the concentration of heavy metal atoms that can diffuse within the FeCoB alloy \cite{devolder_irradiation-induced_2013}.
Despite the strong evolutions of anisotropy and damping, the thickness of Ta has only minor consequences for the magnetic properties sensitive to the degree of structural disorder (Fig.~\ref{METRICS}, middle column): indeed, the ratio of inhomogeneous linewidth broadening $\Delta H_0$ to the effective anisotropy, as well as the ratio of coercivity $H_c$  to effective anisotropy both seem rather independent from the tantalum thickness.  

Regarding the transport properties, tantalum has a weak impact on the RA product and on the TMR ratio (Fig.~\ref{METRICS}, left column) except for 1\r{A} of Ta which leads to low magneto-resistance probably because of insufficient boron-gettering effect. The weak dependence of the transport properties on the other Ta thicknesses is consistent with the fact that the transport properties are essentially determined by the bottom of the bottom FeCoB layer of the FL, i.e. the one in direct contact with the MgO tunneling oxide. The potential damages induced by with the Ta deposition are located 15 \r{A} away from the MgO/FeCoB interface, which is ''far'' in terms of transport through a tunnel barrier \cite{butler_spin-dependent_2001}. Still, the TMR is slightly enhanced by the use of a sacrificial Mg layer, in line with previous findings on MgO/FeCoB/ Mg / Ta single MgO MTJs \cite{swerts_beol_2015}. The damages induced by the direct deposition of Ta are sizable in the 4 \r{A} case: omitting the sacrificial layer leads to a substantial increase of coercivity, and a significant increase of the inhomogeneous linewidth broadening. This detrimental effect is reduced by the insertion of the sacrificial Mg layer before the deposition of the tantalum spacer.

%

In conclusion, we have investigated the properties of MgO/FeCoB/Ta/FeCoB/MgO perpendicularly magnetized free layers embedded in tunnel junctions. After the annealing at 400$^\circ$C, only the samples with 2\r{A} or more of spacer possess perpendicular anisotropy.  A couple of \r{A} of Ta spacer is needed for the top  MgO/FeCoB interface to contribute substantially to the total anisotropy, which is indicative of a better defined top MgO/FeCoB interface. The increase of anisotropy with the Ta thickness is achieved at the expense of a degraded damping that typically increases by 0.001 per added \r{A} of spacer, stating from 0.006. This is likely due to some  interdiffusion between Tantalum and FeCoB occurring during the annealing. A thickness of 3\r{A} of Ta spacer seems the best anisotropy/damping compromise, and its allows both spin-torque operation and thermal stability down to diameters of at least 22 nm. Coercivity and FMR inhomogeneous linewidth broadening can be reduced if a sacrificial Mg layer is inserted before the Ta spacer deposition, meant to reduce the roughness of FeCoB/Ta interface. \\
Acknowledgement: this work was supported in part by the IMEC’s Industrial Affiliation Program on STT-MRAM device, and in part by a public grant overseen by the French National Research Agency (ANR) as part of the Investissements dAvenir Program (Labex NanoSaclay) under Grant ANR-10-LABX-0035.

%

\end{document}